# Influence of Carbon Nanomaterial Counter Electrode Composition, Dye Selection, and Photoanode Scaffolding on DSSC Performance

Benjamin K Barnes[1,2], Joshua Orebiyi[3], Kausik Das[1*]

[1]Department of Natural Sciences, University of Maryland Eastern Shore, Princess Anne, MD 21853, USA

[2]Department of Chemistry and Biochemistry, University of Maryland College Park, MD 20742 USA

[2]School of Engineering, University of Maryland Eastern Shore, Princess Anne, MD 21853, USA

[*]kdas@umes.edu

*Abstract*

As technology continues to evolve, the demand for renewable and sustainable energy continues to grow. As the use of renewable energies, specifically photovoltaics, is continually being adopted and incorporate into everyday life, it is evident that a need for an increase in the amount of energy that is derived from these processes is necessary. In the exploration of dye sensitized solar cells (DSSC) a recent and increased involvement in the application of carbon based nanomaterials and the effects of their unique electronic properties is being investigated. Additionally, as the development of DSSC continues to break way, different methods of dye selection and photoanode scaffolding are being researched to ultimately increase the Power Conversion Efficiency (PCE) of these cells.

*Introduction*

Photovoltaics have received growing attention in the research and industrial setting in recent years due to awareness of the impact that carbon emissions are having on the environment. A recent development in this field is the dye sensitized solar cell (DSSC) which exploits the susceptibility of organic or organometallic dye molecules to visible-light-driven excitation processes; excited electrons are transferred to a simple metal oxide semiconductor and then to a transparent photoanode for use in an external circuit[1]. DSSCs have the benefits of being simple in design, inexpensive, biodegradable, and have the potential to be printed on flexible substrates.

The increased availability of carbon nanomaterials such as nanotubes, nanostars, graphene sheets, and nano fibers has allowed the application of their unique electronic properties to assist in the advancement of DSSCs. Many studies have focused on the incorporation of these and related materials in the photoanode, where they are composited with $TiO_2$ nanoparticles[2,3,4,5,6]. Fewer studies have investigated their use in the catalytic counter electrode however, where electrons are transferred to the oxidized dye by an electrolyte solution, and these studies have been limited to the use of pure graphene, pure carbon black, or pure carbon nanotubes[7,8,9]. Few studies have focused on the use of a blend of these materials with other counter electrode materials such as silver or platinum.

An additional consideration in DSSC design is the identity of the dye compound. The use of simple dyes such as anthocyanins extracted from berries is one attractive feature of DSSCs[10] and much work has investigated the relative performance of different natural dyes extracted from the leaves[11,12,13,14], roots[15], fruits[16,17], flowers[18], and seeds[19,20] of a wide range of plants. While the efficiency of DSSCs based on such natural dyes is generally quite low, these dyes are more stable, less expensive, and often have a higher extinction coefficient than the higher-efficiency organometallic dyes[21].

An additional area of active DSSC research is the photoanode material, design, and morphology, with an emphasis on developing a highly ordered but high surface area structure[22]. One specific improvement in photoanode design is the use of a blocking layer of compact titania particles in order to prevent contact between the electrolyte and anode which would lead to a short circuit or loss of forward current [23]. An additional improvement in photoanode design has been the use of scaffolding layers to tailor the active area thickness. It was shown, for example, that efficiency peaks at around 13 μm of built up P25 titania scaffolds[24].

The goal of this work is firstly to explore the impact of combining a variety of carbon nanomaterials with the catalytic cathode by spreading the particles on the surface of the platinum layer or blending them with the platinum precursor, and secondly to investigate the use of a variety of natural, synthetic, and blended dyes with different types of titania scaffold designs, including different blocking layer and absorbing layer combinations.

*Materials and Methods*

Cathode Experiments

Variation 1: Platisol platinum precursor (Solaronix) was applied to clean FTO glass (Arbor Scientific) with the doctor's blade method. The wet Platisol layer was then sprinkled with Activated carbon (AC from Adrich) or graphene nanostars (GNS from Graphene Supermarket). This was followed by annealing the slide at 450°C for 10 minutes. Photoanodes were made by spin-coating Ti-nanoxide T300/SC (Solaronix) on FTO glass, annealing at 475°C for 30 minutes, and sensitizing with a synthetic ruthenium dye (Ruthenizer 535 bisTBA from Solaronix). The device was completed using binder clips to hold the cathode and photoanode together with a drop of iodide/triiodide-based electrolyte (Mosalyte TDE 250 from Solaronix) in between.

Variation 2: AC or GNS were mixed with the Platisol solution in three different mass ratios (carbon to Platisol), applied to clean FTO glass using the doctor's blade method, and annealed at 450°C for 10 minutes. These were combined with pre-fabricated titania photoanodes (Solaronix) which were sensitized for 10 minutes in Ruthenizer. The device was completed with binder clips and Mosalyte as above.

<u>Photoanode Experiments</u>

Variation 1: Ti-nanoxide BL/SC blocking layer was spin coated onto clean FTO glass and annealed at 475°C for 30 minutes. An absorbing layer of 18 nm $TiO_2$ nanoparticles (Sigma Aldrich) was spin coated on top of this and annealed at 450°C for 30 minutes. These photoanodes were sensitized for 10 minutes in Ruthenizer; for 10 minutes in the pulp of either blackberry, blueberry, or *Syzygium cumini* fruits; or for 5 minutes in Ruthenizer followed by 5 minutes in an equal-volume mixture of the three fruit pulps listed above – in all there were five dye variations. These were combined with plain platinum-coated electrodes (Solaronix). The circuit was completed with Mosalyte as before, and the device was sealed with a Meltonix gasket and Amosil two-part sealing glue (Solaronix).

Variation 2: An absorbing layer of 18 nm $TiO_2$ nanoparticles was spin coated on top of a pre-fabricated $TiO_2$ blocking layer electrode (Solaronix) and annealed at 450°C for 30 minutes. Then was sensitized and combined with a Pt-coated cathode (Solaronix) as described above.

Variation 3: A pre-fabricated $TiO_2$ scaffolding layer electrode (Solaronix) was briefly heated to 200°C to remove residual volatiles and was sensitized and combined with a Pt-coated electrode as described above.

The thicknesses of each layer in these different device designs was measured using a Keyance LK-H022 laser displacement meter and are tabulated below in Figure 3.

The IV behavior of each of the cells was recorded at 1 Sun intensity using an Abet Technologies solar simulator and a Keithley 2450 Source Meter with the 4-wire measurement option.

*Results and Discussion*

<u>Cathode Experiments</u>

The influence of catalytic carbon materials of device performance was first investigated by sprinkling activated carbon or graphene nanostars on a layer of platinum precursor prior to annealing. The control, in this case, was a pre-fabricated platinum-coated electrode (Solaronix). The device design, JV curves, and quantitative performance can be seen below in Figure 1.

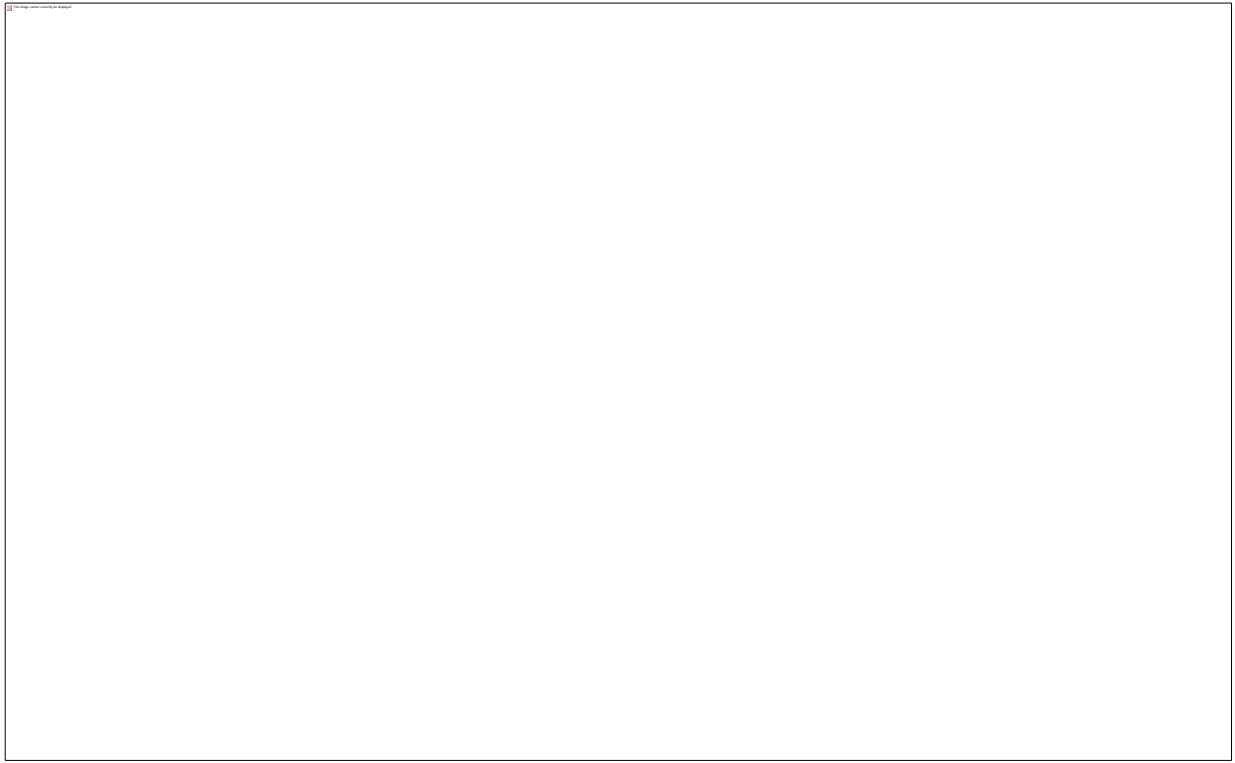

*Figure 1: (Clockwise from top left) Structure of DSSC with carbon particles (activated carbon or graphene nanostars) dispersed on the platinum layer; structure of DSSC with carbon particles mixed with the platinum precursor; JV curve for DSSCs made with a dispersion of carbon particles on the platinum surface; JV curve for DSSCs with carbon particles blended with the platinum precursor; table of PCE and FF values for DSSCs of both designs.*

It is clear from Figure 1 that the addition of carbon nanomaterials to the surface of the Pt layer enhanced the overall performance of the device with respect to the control by increasing power conversion efficiency (PCE) from 0.20% (for the control) to 0.31% and 0.35% for the GNS- and AC-coated cathodes, respectively. Notably, the fill factor (FF) and open circuit voltage ($V_{oc}$) remained roughly equal for all three devices with significant changes only in $I_{sc}$. Variations in $I_{sc}$ for DSSCs made with carbon nanomaterial catalysts are typically attributed to the surface area and purity of the carbon material used[25,26].

Next, we sought to investigate the effect of blending carbon nanomaterials directly into the platinum precursor solution prior to coating the FTO glass. For both GNS and AC blended with Platisol, efficiency was significantly decreased when compared with the previous experiment wherein the same materials were simply deposited on the surface of the platinum precursor. The layer of carbon material deposited on the surface of the Platisol layer in the previous experiment likely resulted in better contact with the electrolyte solution than when blended into the Platisol layer, hence the increased efficiency. In this case, the nanostars performed significantly better than did activated carbon, exhibiting larger PCE, $V_{oc}$, and $I_{sc}$ values. Increasing the AC content in fact decreased PCE and resulted in a breakdown of the JV curve by around 4% (w/w). This may be due to a loss of the carbon-based material at the high-temperature annealing step which would lead to voids in the catalyst layer. Increased GNS content, however, increased the efficiency of the cell but had little impact on the FF.

Photoanode Experiments

Next, a variety of photoanode designs were evaluated with different dyes. The first photoanode design consisted of a Ti-nanoxide BL/SC blocking layer followed by a $TiO_2$ nanoparticle light absorbing layer combined with a plain Pt counter electrode. This device design and the associated JV curve can be seen in Figure 2 and Figure 3 below. The cell sensitized with the combination of mixed fruit dyes and Ruthenizer exhibited the best performance, with a PCE of 0.10% and FF of 0.50. The three cells using pure fruit pulp dyes exhibited much lower PCE, $V_{oc}$ and $I_{sc}$ values. However, it should be noted that the FF of the cell made with *S. cumini* dye was about 0.52 in contrast with the much lower values of the blackberry- and blueberry-sensitized cells (0.40 and 0.38, respectively). It should be noted that the stain made on the photoanode by *S. cumini* pulp was much lighter in color than the other dyes used, indicating incomplete dye absorption; this has been correlated to increased FF values by in some studies[27], and may explain the difference in observed FF here.

Next, the cells made by depositing a film of $TiO_2$ nanoparticles on the pre-fabricated blocking layer were investigated. The cell structure and associated JV curves can also be seen below in Figure 2 and Figure 3. The most striking result shown in this set of plots is the enhanced FF when compared to that of the cells discussed above; the FF values for all of these cells made with pre-fabricate blocking layers were around 0.60. The low FF values associated with the previous data set indicate that the device design used there was subject to parasitic resistive forces not present in this design[28]. It is also notable that the cell in this set with the highest PCE value (0.07%) was the one sensitized with blackberry pulp, and not the mixture of dyes as it was in the previous set; though the two PCE values for these two cells are comparable.

The next photoanode design to be evaluated was the pre-fabricated scaffolding layer which consisted of a pre-fabricated absorbing layer deposited on a pre-fabricated blocking layer. The structure of this device and its associated JV curves for different dyes can be seen below in Figure 2 and Figure 3. In this case, the PCE behavior was quite similar to that of the "Ti-nanoxide BL/SC and nanoparticle" design discussed above, with the mixed dye performing very well and all other berry dyes performing less well but similarly. Again, the FF of these devices was much higher than those of the first set discussed, and ranged in value between 0.54 and 0.66.

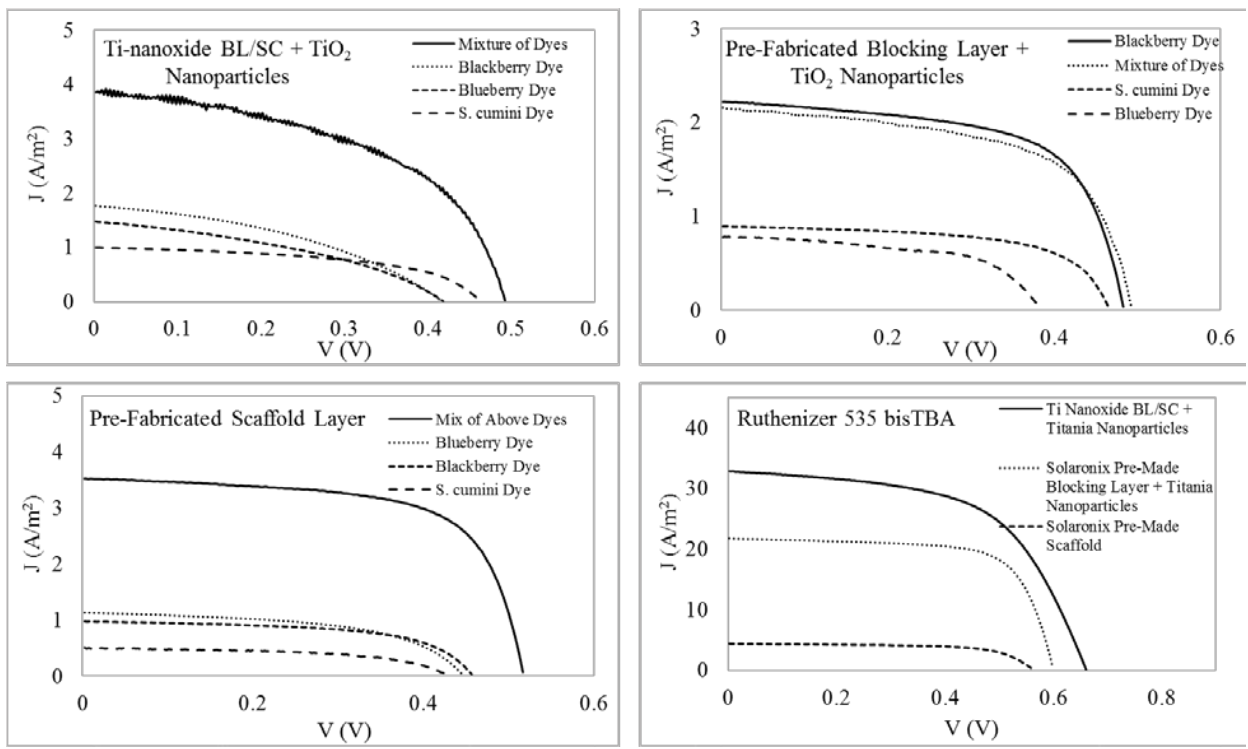

*Figure 2: (Clockwise from top left) JV curve of DSSCs fabricated with TiO₂ nanoparticles deposited on a Ti-nanoxide BL/SC blocking layer; JV curve of DSSCs fabricated with TiO₂ nanoparticles deposited on a pre-fabricated blocking layer; JV curve of DSSCs of three different photoanode designs sensitized with Ruthenizer; JV curve of DSSCs made from pre-fabricated scaffolding layer photoanodes.*

Finally, each photoanode design was also evaluated using only pure Ruthenizer dye solution; previously, berry pulp or a combination of berry pulp and Ruthenizer was utilized. As expected, these cells yielded much higher PCE values than did the same designs sensitized with fruit pulp. It was unexpected however, that they would out-perform the cells made with a mixture of Ruthenizer and fruit juice, as previous results had indicated an additive effect of blending the dye types, but this was not the case for each photoanode variation in this experiment. This may have resulted from different sensitization times, as the mixed-dye DSSCs were sensitized in Ruthenizer for 5 minutes followed by 5 minutes in the berry mixture, whereas the Ruthenizer cell was sensitized for 10 minutes in Ruthenizer. As previous work has demonstrated that Fick diffusion is observed in DSSC sensitization[27,29], it is likely that the higher-performing Ruthenizer failed to penetrate sufficiently for a soak time of 5 minutes in the case of the mixed dye treatment, whereas it penetrated fully in the case of pure Ruthenizer sensitization. Interestingly, the FF trend for the cell made with Ti-nanoxide BL/SC and nanoparticles held true, and was significantly lower than the values for the other two designs (0.57 as opposed to 0.71 and 0.67). The same device, however, yielded the highest PCE at 1.24% compared with 0.92% and 0.16% for the other two designs.

A summary of all three photoanode designs and the associated thicknesses of the various layers can be seen below in Figure 3. Also shown is the UV-visible absorption spectrum of the different dyes used; the strong absorption of Ruthenizer at around 420 nm and 520 nm is likely a source of this dye's excellent performance in this work, as the natural dyes from the fruit pulp lacked strong absorption in these regions.

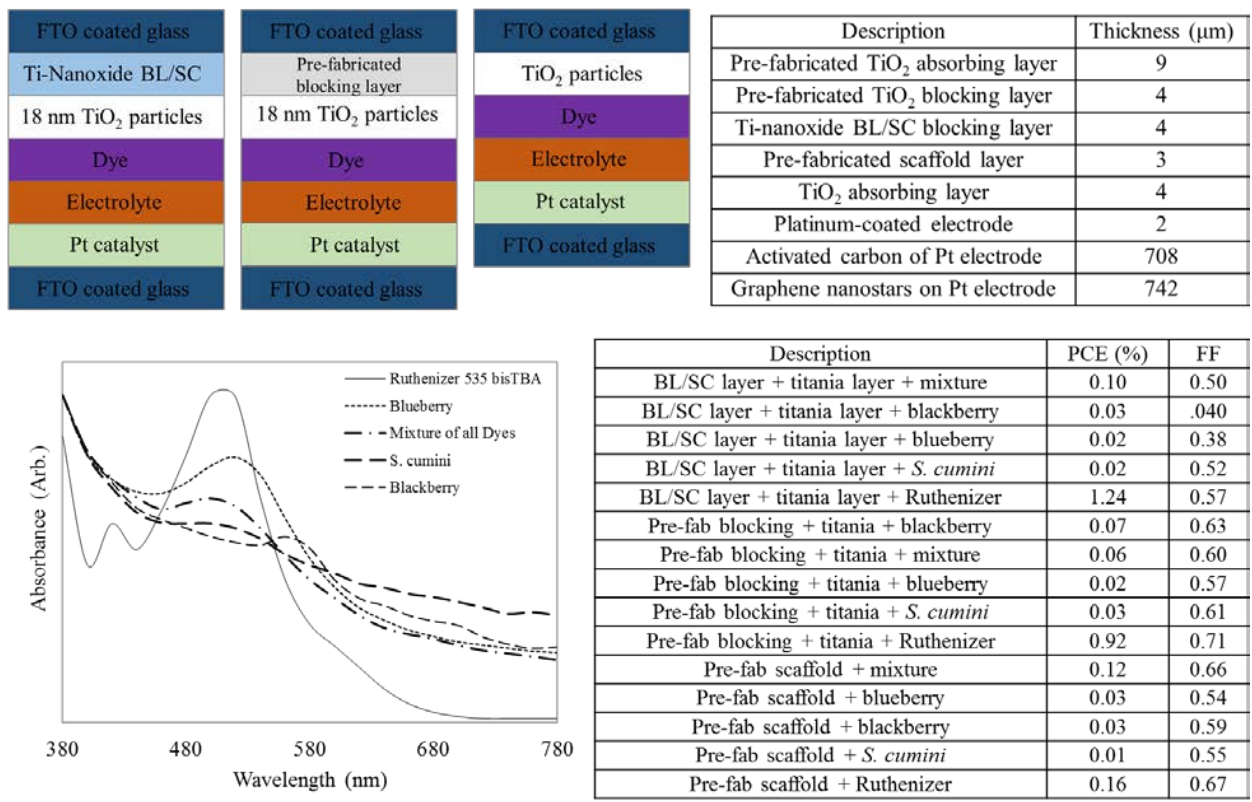

*Figure 3: (Clockwise from top left) Three different photoanode designs, each sensitized with five different dye combinations; measured thickness of the different layers in the DSSC designs; tabulated PCE and FF values of all DSSCs from the photoanode experiments; absorption spectra of the various dye combinations used.*

*Conclusion*

In this work, a wide variety of DSSCs were fabricated and evaluated in order to explore the result of modifying the catalytic materials, photoanode design, and sensitizing dyes. It was found that applying a thin layer of carbon nanomaterials (activated carbon and graphene) enhanced the performance of the DSSC by increasing the short circuit current. It was also found that mixing the same carbon particles into the platinum precursor significantly reduced efficiency; this may be the result of the pyrolysis of carbon particles leaving voids in the catalytic layer, but more work is needed to verify this. Three variations in photoanode structure were also evaluated with five different dyes each. The highest PCE observed was 1.24%; this value was attained using a photoanode design consisting of a Ti-nanoxide BL/SC blocking layer and an absorbing layer of 18nm $TiO_2$ nanoparticles sensitized with Ruthenizer 535 bisTBA. This same DSSC, however, had a much lower FF value than did the other designs, a trend which held true for all dye species used. The second best performance was observed in a similar cell, but one which was made with a pre-fabricated blocking layer; the PCE of this cell was 0.92%, and at 0.71, it had the highest FF value of all the DSSCs tested. DSSCs of all photoanode designs performed well when sensitized with a mixture of dyes (from three different fruit pulps and Ruthenizer), though incomplete sensitization by Ruthenizer seems to have limited the PCE of these cells.

*Acknowledgement*

This research was supported by National Science Foundation HBCU Up grant # 1719425 and Department of Education MSEIP grant # P120A170068.